# Chemokinesis-Driven Accumulation of Active Colloids in Low-Mobility Regions of Fuel Gradients

Jeffrey L. Moran*[1][‡], Philip M. Wheat[2][‡], Nathan A. Marine[2], & Jonathan D. Posner[3,4,5]

[1]Department of Mechanical Engineering, George Mason University, Fairfax, VA

[2]Ira A. Fulton Schools of Engineering, Arizona State University, Tempe, AZ

[3]Department of Mechanical Engineering, University of Washington, Seattle, WA

[4]Department of Chemical Engineering, University of Washington, Seattle, WA

[5]Department of Family Medicine, School of Medicine, University of Washington, Seattle, WA

([‡] These authors contributed equally)

(* Corresponding author)

## Abstract

Many motile cells exhibit migratory behaviors, such as chemotaxis (motion up or down a chemical gradient) or chemokinesis (dependence of speed on chemical concentration), which enable them to carry out vital functions including immune response, egg fertilization, and predator evasion. These have inspired researchers to develop self-propelled colloidal analogues to biological microswimmers, known as active colloids, that perform similar feats. Here, we study the behavior of half-platinum half-gold (Pt/Au) self-propelled rods in antiparallel gradients of hydrogen peroxide fuel and salt, which tend to increase and decrease the rods' speed, respectively. Brownian Dynamics simulations, a Fokker-Planck theoretical model, and experiments demonstrate that, at steady state, the rods accumulate in low-speed (salt-rich, peroxide-poor) regions not because of chemotaxis, but because of chemokinesis. Chemokinesis is distinct from chemotaxis in that no directional sensing or reorientation capabilities are required. The agreement between simulations, model, and experiments bolsters the role of chemokinesis in this system. This work suggests a novel strategy of exploiting chemokinesis to effect accumulation of motile colloids in desired areas.

## I.1 Chemotaxis vs. Chemokinesis

Self-propulsion at the microscale is abundant and important in biology. Many eukaryotic and prokaryotic cells propel themselves in biological fluids to perform critical functions. Examples include sperm cells, which navigate through cervical mucus to fuse with and fertilize an egg[1], and leukocytes, whose migration is essential for coordinated immune responses and often occurs over long distances[2,3]. Two common migratory behaviors exhibited by motile cells are *taxis* and *kinesis*. Taxis refers to the phenomenon in which the *direction* of an organism's motion is determined by the non-uniform distribution of a physical quantity (i.e., the cell moves up or down a gradient). In contrast, kinesis occurs when an organism's *speed*, either translational or rotational, depends on the spatial distribution of a quantity[2]. Unlike taxis, kinesis does not imply a direction and is generally random.

In the natural world, taxis manifests in many forms, including chemotaxis (migration up or down a chemical concentration gradient), aerotaxis (oxygen gradient), phototaxis (light intensity gradient), gravitaxis (gravitational potential gradient), and others.[4–6] Taxis drives many cell migration behaviors known to be essential for various physiological and pathological processes. For example, motile bacteria execute chemotaxis to find nutrients and evade predators[7]. Aerobic bacteria perform aerotaxis to find the oxygen they need to survive[8]. Leukocytes reach sites of infection by orienting toward higher concentrations of chemicals secreted at these sites[2,9].

To perform taxis, cells employ a coordinated series of sensing and signaling processes. For example, *Escherichia coli* bacteria possess transmembrane receptors that bind attractants (such as glucose) and repellents (such as phenol) and thereby detect spatial or temporal differences in chemical concentration. These differences are relayed via a signaling protein to the flagella, which execute runs (straight-line motions) and tumbles (i.e., orientation changes), adjusting the cell's direction according to its needs. In addition to spatial gradients, bacteria can also respond to temporal gradients. For example, *Salmonella typhimurium* bacteria change their tumbling frequency, effectively modifying direction, in response to a change from one uniform concentration of a chemoattractant to another[10]. This suggests that the detection of a spatial gradient by bacteria may involve detection of a temporal concentration variation, which the cell experiences by moving through a spatial concentration difference. In contrast, eukaryotic cells (such as leukocytes) detect spatial gradients by comparing the occupancy of receptors at different locations along the cell[11]. Although the details differ across cell types, taxis generally requires sensing, signal transduction, and movement response.[12]

In contrast to taxis, kinesis implies that cells or organisms move at a speed that depends on a local stimulus intensity. One common form of kinesis is chemokinesis, in which the organism's speed depends on chemical concentration. The dependence of translational speed on stimulus is known as *orthokinesis* and dependence of rotational speed or reorientation frequency on stimulus is known as *klinokinesis*. In a concentration gradient, cells undergoing chemokinesis exhibit a spatial variation in motility[2], which can result in accumulation of cells in high- or low-concentration regions[13]. Chemokinesis has been observed in spermatozoa[14,15], neural cells[16], leukocytes[17], and bacteria such as *Rhodobacter sphaeroides*[18]. Numerical models of simultaneous chemotaxis and chemokinesis, based on the behavior of *Myxococcus xanthus*, demonstrate how the two behaviors can coexist competitively in an organism[19]. In a chemical gradient, chemokinesis can lead to accumulation in regions where motility is minimized. For example, neutrophils swim more slowly in the presence of an immune complex than in the presence of surfaces coated with bovine serum albumin; over time, the neutrophils were shown to accumulate preferentially in the immune complex region[20].

In summary, when cells are exposed to chemical gradients, chemokinesis and chemotaxis can both arise. Perhaps as a result, there has been confusion in the literature between chemotaxis and chemokinesis.[9,13] In a bounded domain with a static concentration gradient, chemokinesis can result in accumulation of cells that appears qualitatively similar to, and may be mistaken for, chemotaxis[9,13]. The key distinction is that, in chemokinesis, cells can accumulate in certain regions not by sensing and "intentionally" responding to a gradient, but because of the coupling between stochastic variations of their motion direction (e.g. due to tumbling or rotational Brownian motion), which allows cells to explore their surroundings and sample the surrounding concentration field, and the kinetic response, which causes speeds to decrease in certain areas (effectively trapping cells there)[13,20].

## I.2 Self-propelled microparticles: vehicles for artificial chemotaxis?

Since the early 2000s, extensive research efforts have been devoted to developing synthetic analogues to biological microswimmers. These self-propelled particles, often called *active colloids*, typically range in size from 0.1 to 10 μm and have the unique ability to harvest energy from their surroundings and convert it to motion[21]. They accomplish this via several mechanisms including bubble propulsion[22], conversion of external electric or magnetic fields[23,24], ultrasound[25,26], or incident light[27] into motion, and phoretic mechanisms[28] in which the particle moves as a result of self-generated gradients in temperature, chemical concentration, or electric potential. The first active colloids to be widely studied were half-platinum half-gold (Pt/Au) bimetallic rods (2 μm long, between 200-400 nm diameter) that self-propel with the Pt end forward in hydrogen peroxide ($H_2O_2$) solutions.[29] Several mechanistic analyses[30,31], including some from our group[32–34], established that electrochemical reactions on the Pt and Au surfaces generate an electric field in the rod's vicinity, which exerts a propulsive force on the charged rods (commonly called self-electrophoresis[30,32,35,36]). Figure 1 shows a schematic of a Pt/Au rod with the electrochemical reactions powering the motion.

The speed of Pt/Au rods increases with local $H_2O_2$ concentration, and the rods thus exhibit a positive chemokinetic response to $H_2O_2$. Some studies report a linear dependence of speed on concentration while others predict nonlinear relationships of varying types[29,32,33,37,38]. The general consensus is that the dependence is linear at low to moderate concentrations, and slightly levels off at high concentrations because of saturation of available reaction sites on the Pt and Au surfaces[37,38]. Other active colloid designs also exhibit a chemokinetic response to their fuel (typically $H_2O_2$), including bubble-propelled particles[39] and Pt/polystyrene or Pt/silica "Janus" particles[40,41]. At higher fuel concentrations, speeds are usually faster because the reactions responsible for motion proceed more rapidly, leading to stronger propulsive forces.

In contrast to $H_2O_2$, the addition of electrolytes tends to decrease the speed of Pt/Au rods. This was first reported in 2006 for sodium nitrate and lithium nitrate by Paxton et al.[42], who found that at constant $H_2O_2$ concentration, speed is roughly inversely proportional to solution conductivity. ($H_2O_2$ is required for electrolyte-induced speed reductions to be present, since without H2O2 there would be no self-propelled motion to impede.) The only exception to this trend is electrolytes that contain silver, which tend to *increase* the speed of the rods[43]. Electrolytes not containing silver cause significant reductions in rod speed even at modest concentration, such that millimolar concentrations are often enough to effectively eliminate self-propulsion[42]. To elucidate the physical mechanism for this phenomenon, we simulated the motion of a single rod in varying salt solutions[34] and found that the self-generated electric field magnitude decreases markedly upon addition of salt (and the concomitant increase in conductivity). This trend is robust and has been characterized for a wide range of electrolytes[43].

Several groups have attempted to realize autonomous taxis in artificial active colloids[44]. Some studies[45–48] have reported chemotaxis in chemically-powered particles, effectively making two simultaneous claims: (1) the particles move by consuming a chemical fuel that influences their speed (chemokinesis); (2) the particles move up gradients of that same fuel (chemotaxis). As previously stated, chemotaxis and chemokinesis can coexist in some organisms[19]; however, given the signaling and sensing required for chemotaxis in living cells[49], the question naturally arises as to how synthetic active colloids could achieve "intentional" motion along gradients. The argument is often made that when an active colloid happens to move up a fuel gradient (i.e., in a direction of increasing concentration), it accelerates and therefore advances farther toward increasing concentration than if it were moving toward lower concentration. This scenario is plausible but does not address the scenario in which a particle encounters a local maximum in chemical concentration. When a chemokinetic active colloid reaches a concentration maximum, it will quickly disperse because of the maximal motility there. In contrast, chemotactic cells do not disperse upon reaching a chemoattractant source; rather, their directional motion ceases. For example, chemotactic bacteria seek out and accumulate at nutrient maxima[50]. Sperm cells accumulate in follicular fluid containing a chemoattractant[51], and there is a strong correlation between this *in vitro* accumulation and egg fertilizability[52]. Neutrophils use chemotaxis to find and accumulate at wound sites[53]. A recent numerical study considered a self-propelled particle in a linear fuel gradient and assumed a linear dependence of speed on chemical concentration; the main conclusion was that, although initially the particle may move in either direction, it would ultimately end up in the region of lowest motility[55]. As noted by Popescu et al. in a systematic study of Janus swimmers in concentration gradients[54], "chemokinesis alone cannot lead to chemotaxis."

Some theoretical analyses have been undertaken of phoretic active colloids to identify the conditions in which they could exhibit chemotaxis, positing a rich variety of single-particle and collective behaviors[56,57]. It is well-known from the work of Golestanian and others that the response of a phoretic self-propelled particle depends on its phoretic mobility and surface activity[35]. Popescu et al. theoretically analyzed the response of a phoretic self-propelled particle to a concentration gradient and concluded that, while a particle's chemokinetic response to a fuel depends primarily on its *average* phoretic mobility (in the case of Pt/Au rods, this is the electrophoretic mobility, which depends on zeta potential[58]), *chemotaxis* primarily depends on the mismatch in the phoretic mobility between the front and back halves of the particle[54]. The zeta potential of metallic nanorod particles in aqueous solutions has been measured to be consistently negative (attributed to preferential adsorption of anions)[59], and is around –40 mV for Pt/Au rods[42]. Since the particle is conducting, our previous theoretical analyses have predicted that the zeta potential of metallic nanorods differs negligibly between the Pt and Au surfaces[33]. Thus, the theoretical basis for Pt/Au active colloids to exhibit chemotaxis of this type has not been definitively established.

Other demonstrations of artificial taxis are based on different physical mechanisms. Silica/carbon two-faced "Janus" microspheres, when immersed in a binary mixture of water and 2,6-lutidine and irradiated with 532-nm light, asymmetrically de-mix the solution surrounding the particle, establishing a lutidine concentration gradient that produces a diffusiophoretic propulsive force on the particle[60]. In a light intensity gradient, a torque on the particle arises that selectively orients particles toward low-intensity regions, exemplifying artificial negative phototaxis. In another example, droplets of a liquid crystal move by generating surfactant gradients, driving Marangoni flows that propel the droplets[61]. The droplets evinced chemotaxis by successfully navigating a microfluidic maze with the chemoattractant source at the exit. Other examples include asymmetric artificial liposomes made from copolymer mixtures, which undergo runs and tumbles akin to bacteria and are propelled by enzymatic reactions[62]; these particles were shown to be promising for blood-brain barrier crossing, a crucial step in many biomedical applications. Finally, gravitaxis was demonstrated in asymmetrically-coated Janus microspheres[63]; here, the mismatch between the densities of the polystyrene colloids and metallic coatings led to a torque orienting the particles along the gravitational direction with the metallic side down.

In contrast to taxis, a lesser-explored strategy is to exploit the chemokinetic response that most active colloids naturally exhibit to bring about their accumulation in desired areas. One recent example study[64] demonstrated polymeric self-propelled particles that swell in response to decreases in pH. In a pH gradient, particles accumulate in the most acidic regions because the PVP swells in response to the local pH decrease, leading to an increased drag profile and reduced swimming speed there. Given that many active colloids naturally exhibit chemokinesis, a systematic characterization of the behavior of active colloids in static concentration gradients is warranted.

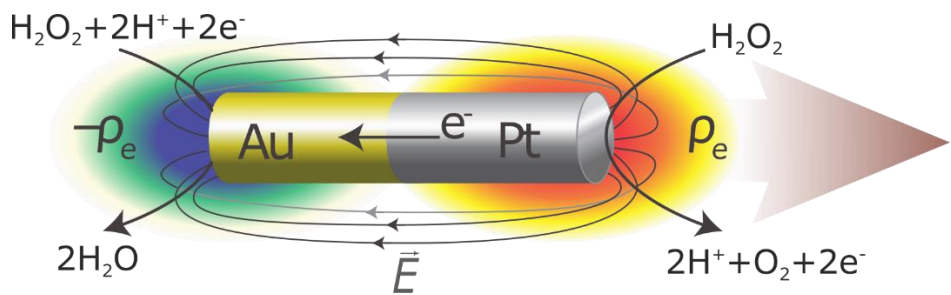

**Figure 1: Schematic of self-propelled platinum/gold (Pt/Au) nanorods used in this work.** First reported in 2004[29], Pt/Au nanorods move autonomously with the Pt end forward in aqueous hydrogen peroxide ($H_2O_2$) solutions. Pt and Au catalyze the indicated electrochemical reactions, resulting in an electron current through the rod (from Pt to Au) and a corresponding ion current in the solution, and create dipolar charge density ($\rho_e$) and electric field (E) distributions around the rod. The electric field exerts a propulsive force on the negatively-charged rod, propelling it with the Pt end forward (left to right)[32,33]. The speed of Pt/Au rods increases as the $H_2O_2$ concentration is increased, while the addition of a nonreacting electrolyte tends to decrease the self-propelled speed[34,42].

In this paper, we show that Pt/Au rods undergo chemokinesis-driven accumulation in steady-state concentration gradients of hydrogen peroxide ($H_2O_2$) and potassium chloride (KCl) salt. We use a microfluidic device to generate steady-state linear gradients of $H_2O_2$ and KCl. The rods are initially distributed randomly in these gradients and exhibit a positive and negative orthokinetic response to $H_2O_2$ and KCl, respectively. The distribution of rods reaches a pseudo-equilibrium and concentrates in regions with low $H_2O_2$ and high KCl concentrations, where their effective diffusivity is minimized. The experiments show good agreement with Brownian dynamics simulations as well as a theoretical model based on a 1-D Fokker-Planck equation, which models the particles as a continuous substance undergoing enhanced diffusion with spatially-dependent diffusivity. The simulations, PDE model, and experiments point to a simple explanation that chemokinesis can lead to accumulation of active colloids in chemical gradients. This accumulation is distinct from chemotaxis in that it does not require sensing of, or a direct response to, temporal or spatial gradients in concentration.

II.1 Methods

The experiments were conducted in a microfluidic device fabricated from polycarbonate, glass, double-adhesive Mylar, and nitrocellulose membranes (see Figure 2a). All layers are patterned using a $CO_2$ laser ablator (Universal Laser Systems, Scottsdale, AZ). The lower and upper surfaces of the microchannels are formed from a glass microscope slide and poly(methyl methacrylate) (PMMA) sheets, respectively. Nitrocellulose membranes (Whatman, Maidstone, UK), with average pore diameter of 0.4 μm, form the boundaries between the main channel and the side channels and are sealed to the upper and lower channel surfaces using 200-μm-thick double sided adhesive Mylar sheets cut in the same pattern as the membranes. The device is assembled using metal bolts through the acrylic upper surface and an additional PMMA superstructure below the glass lower surface. The center microfluidic channel is 15 mm long, 400 μm wide, and 450 μm deep. The outer channels are 600 μm wide and 450 μm deep. The assembled device is depicted schematically in Figure 2a.

The rods are fabricated using a templated electrodeposition procedure[65] that is widely used for synthesis of self-propelled rod-shaped particles[29,31,32]. Briefly, platinum and gold are sequentially electrodeposited into the cylindrical 200-nm-diameter pores of an anodic aluminum oxide (AAO) membrane. After metal electrodeposition is complete, the AAO membrane is chemically etched, and the rods are centrifuged and resuspended in pure water. The length of the Pt and Au segments is each approximately 1 μm, yielding an overall rod length of 2 μm. The rods are imaged in the microfluidic device using optical microscopy with a 20× objective (NA = 0.45, Nikon TE2000, Japan) and a CCD camera (Coolsnap HQ, Photometric, Tucson, AZ) at 2 frames per second. The motion of the rods is tracked and analyzed using a custom MATLAB-based particle tracking algorithm. In each frame, the positions of the rods' centers were calculated from the intensity-weighted centers of the rod images. The rods' centers at each time were paired using an optical flow algorithm. In a typical experiment, there are between 350 and 600 rods in the field of view. The width of the channel is divided into 20 segments. The mean squared displacement (MSD) for each rod is tracked for 100 frames. An effective diffusivity for each of the 20 segments is determined by averaging the MSDs for all rods starting in the respective segments at the start of the 100 frames.

We drive flow of 30% $H_2O_2$ in deionized water at 5 μL/min through the left channel, which serves as an $H_2O_2$ source and KCl sink. We drive 100 μM KCl + 1 μM fluorescein dye (the dye is used to visualize the concentration gradient using optical microscopy) through the right channel, which serves as a KCl source and $H_2O_2$ sink. The flows in the side channels are driven with syringe pumps. As described in the supplementary information, the maximum and minimum concentrations of KCl and fluorescein in the main channel are calculated as $C_{min,KCl}$ = 45.5 μM, and $C_{max,KCl}$ = 54.5 μM. $C_{min,Fl}$ = 455 nM, and $C_{max,Fl}$ = 545 nM. As shown in Figure 2b, the KCl concentration distribution (yellow solid line) is verified based on the observed variation of fluorescein intensity and assuming the KCl concentration to be linearly proportional to fluorescein concentration. The $H_2O_2$ concentration distribution (red dashed line) is estimated based on the steady-state diffusion equation. The center channel contains the Pt/Au rods in deionized water with no imposed bulk flow. The porous microchannel walls and lack of flow in the center channel result in linear static gradients of $H_2O_2$ and KCl. Pt/Au rods exhibit orthokinetic behavior wherein the self-propelled translational speed depends on local peroxide and electrolyte concentration.

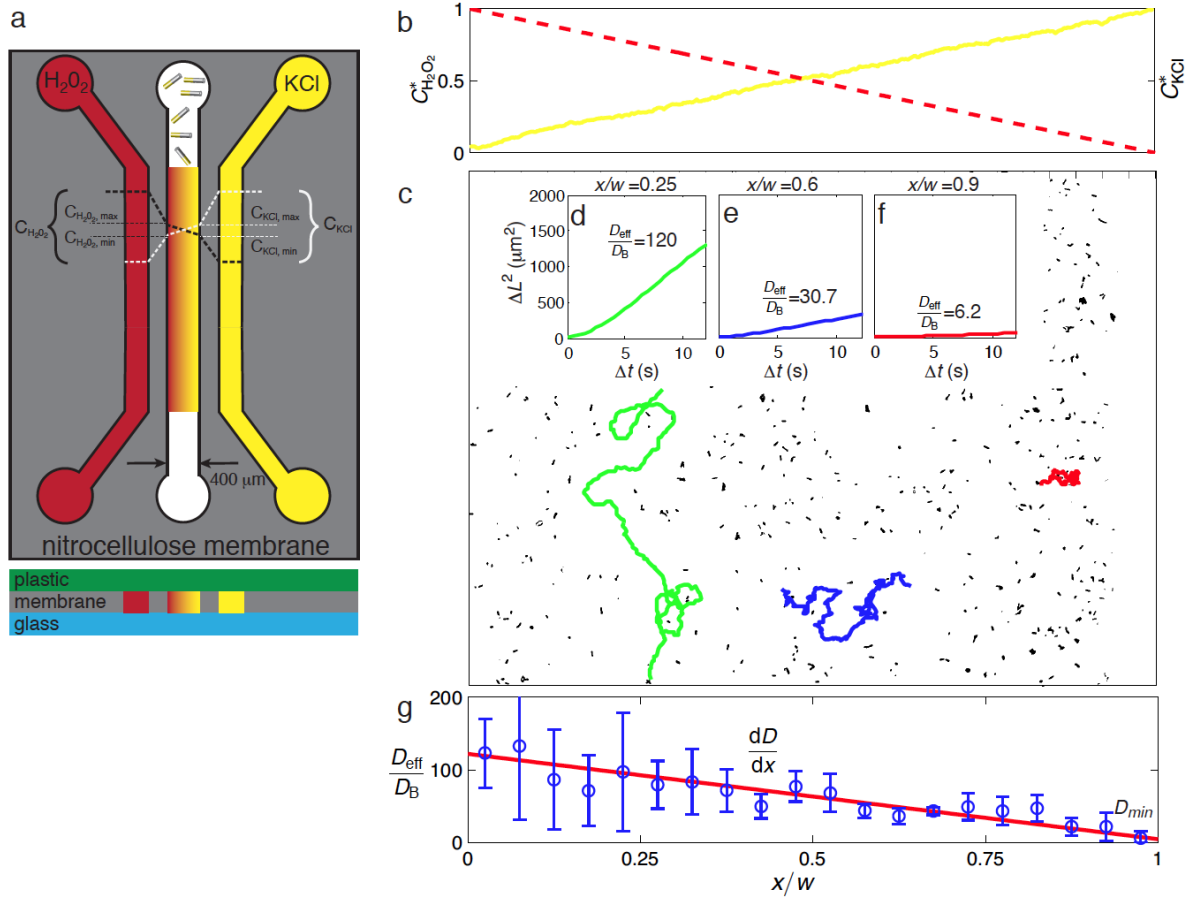

**Figure 2: Microfluidic linear gradient generator and its effect on effective diffusivity of bimetallic rods.** (a) A microfluidic device with porous nitrocellulose channel walls generates steady, linear gradients of hydrogen peroxide ($H_2O_2$) and potassium chloride (KCl) in the center channel. The left (red) channel contains 30 wt.% $H_2O_2$ and the right (yellow) channel contains 100 μM KCl in water. The left and right channels serve as reservoirs for $H_2O_2$ and KCl respectively. The rods are introduced into the center channel, where they move autonomously in the static fluid and at a speed that depends on the local chemical concentrations. (b) Linear gradients of KCl (measured estimates) and $H_2O_2$ (calculated) in the center channel where $C_i^*(x) = (C_i(x) - C_{i,min})/(C_{i,max} - C_{i,min})$. Here x denotes the horizontal spatial coordinate and i denotes either $H_2O_2$ or KCl. (c) Trajectories of individual rods in the center channel in the presence of the $H_2O_2$ and KCl gradients over 100 seconds. The linear static gradients result in a spatial variation in rod swimming speed that results in longer paths on the left (green) relative to the ones on the right (red) due to the locally high $H_2O_2$ concentration and low KCl concentration. Ensemble-averaged mean-square displacement (MSD) as function of time for (d) rods centered at x/w=0.25 (w = 400 μm is the width of the channel) and $D_{eff}/D_B$ = 120, where $D_B$ = 1 μm$^2$/s is the Stokes-Einstein-Sutherland diffusivity of the rod; (e) rods centered at x/w = 0.6 ($D_{eff}/D_B$ = 30.7); and (f) rods centered at x/w = 0.9 ($D_{eff}/D_B$ = 6.2). (g) The measured effective diffusivity (blue circles) varies roughly linearly with x. The error bars represent one standard deviation of the measured diffusivities and the red solid line is a least-squares linear fit that is used in the models. Panels (c)-(g) originally appeared in a previous review by our group[28].

### II.2 Theory

Complementing the experiments, we model chemokinesis using Brownian dynamics (BD) simulations as well as a mass conservation partial differential equation (PDE). The BD simulations are based on the standard Langevin model that neglects inertial effects, collisions between rods, and interparticle hydrodynamic and electrostatic interactions. We assume the motion of the rods is confined to two dimensions, since the negatively buoyant rods sediment to a quasi-2-D fluid layer a few micrometers above the bottom surface, where they remain for the duration of the experiment. Many of the materials from which active colloids are made are denser than water, and thus 2-D (or even 1-D) models of ensembles of active colloids are common[40,55,66–68].

The differential equations governing the position $x$ and $y$ and orientation $\theta$ of each rod are the Langevin equations[55,66,67,69],

$$\dot{x} = u \cos \theta + \xi_x, \tag{1}$$

$$\dot{y} = u \sin \theta + \xi_y, \tag{2}$$

$$\dot{\theta} = \xi_\theta, \tag{3}$$

where $\dot{x}$ and $\dot{y}$ are (respectively) the translational speeds in the $x$ and $y$ directions, $u$ is the translational speed of the rod (which depends on $x$), and $\xi_x$, $\xi_y$, and $\xi_\theta$ are stochastic noise terms in the indicated directions. This basic system of Langevin equations has been used to model quasi-2-D swarms of chemokinetic active colloids in several previous studies[55,66–68,70,71]. The right-hand side of the third Langevin equation (3) includes only the stochastic term, $\xi_\theta$, reflecting our assumption that the rods are not "self-polarizing[66]." In other words, we assume that the orientation of the rods varies in time because of rotational diffusion alone, and that the rotational diffusivity is approximately independent of $H_2O_2$ or KCl concentration. The assumption of no self-polarization has been made in several previous models of chemokinetic microswimmers[40,55,70,72]. For many phoretic self-propelled particles that show a strong chemokinetic response, the rotational diffusivity is primarily governed by rotational Brownian motion and does not depend strongly on stimulus[67,72,73]. For Janus active colloids whose front and back mobilities are nearly equal, the systematic analysis by Popescu et al. found that the reorienting torque, and thus the chemotactic response, are negligible[54].

The BD simulations do not specify an explicit dependence of speed on either $H_2O_2$ or KCl concentration. As discussed above, the dependence of speed on both concentrations has been measured by several groups, including ours[29,32,37,42,43]. Although the qualitative trends are well-established (speed increases approximately linearly with H2O2 concentration and with the reciprocal of KCl concentration), different studies report different values of the rod speed for the same $H_2O_2$ concentration[29,32,37]. In this work, we found that even the same batch of rods showed different average speeds from one day to another under nominally identical conditions. In experiments, ensemble averages of rod particle tracking showed a roughly linear dependence of rod speed on horizontal position $x$[74]. As a result, in the BD simulations presented here, we assumed a linear variation of $u$ with $x$, bounded by maximum and minimum speeds, $V_{max}$ and $V_{min}$, which were specified inputs for each simulation run. The minimum value of $V_{min}$ we considered was 1 μm/s and the maximum value of $V_{max}$ was 25 μm/s. This is a similar range to the speeds measured in the experiments[74].

The stochastic displacements $\xi_x$, $\xi_y$, and $\xi_\theta$ are independent random variables with zero ensemble average, $\langle \xi_x \rangle = \langle \xi_y \rangle = \langle \xi_\theta \rangle = 0$, and their standard deviations are tuned such that the mean-squared displacements (without self-propulsion) are consistent with Brownian motion with translational diffusivity $D$ and rotational diffusivity $D_{rot}$,

$$\langle \xi_x^2 \rangle = \langle \xi_y^2 \rangle = 2Dt, \tag{4}$$

$$\langle S_\theta^2 \rangle = 2D_{\text{rot}}t, \tag{5}$$

where brackets indicate an ensemble average over the population of rods, $D$ is the Brownian diffusion coefficient of the rods, and $D_{rot}$ is the Brownian rotational diffusivity of the rods. For the purposes of tuning the noise terms, the translational and rotational Brownian diffusivities are estimated from the Stokes-Einstein equations for translational and rotational diffusion of a spheroidal particle[75]. For the dimensions considered here, $D \approx 1$ μm$^2$/s and $D_{rot} \approx 1.38$ rad$^2$/s. The boundary conditions employed for the BD simulations enforce no flux of rods (i.e., the channel is closed and rods cannot escape) and elastic boundaries. That is, rods are assumed to undergo perfectly elastic collisions with the walls and the overall kinetic energy of the rod is the same before and after the collision. The simulations track 10,000 rods over 6,000 s with time step $\Delta t = 0.1$ s.

The BD simulations and experiments are complemented with a theoretical model. This model assumes the swarm of rods to be a continuously distributed substance that obeys the mass conservation equation

$$\frac{\partial n}{\partial t} + \nabla \cdot \boldsymbol{J} = 0, \tag{6}$$

where $n$ is the number density of rods and the flux of rods $\boldsymbol{J}$ is defined by the modified Fokker-Planck equation[76–78],

$$\boldsymbol{J} = -D_{eff}\nabla n - \alpha n \, \nabla D_{eff}, \tag{7}$$

where $D_{eff}$ is the effective diffusivity of the rods (which depends, in turn, on the spatially-dependent rod speed), $n$ is the number density of the rods, and $\alpha$ represents the Itô-Stratonovich convention coefficient. The Fokker-Planck equation has been used to model chemokinesis of active colloids in activation gradients in a variety of situations[55,66,67]. The supporting information (SI) provides a full derivation of equation (2). The appropriate value of $\alpha$ for a random-walk process is 0.5 as discussed by in detail by Schnitzer[76]. Equation (6) is solved in one dimension with no-flux boundary conditions, $\boldsymbol{J} = \boldsymbol{0}$, i.e. $D_{\text{eff}}\nabla n = -\alpha n \nabla D_{\text{eff}}$. The only input into the models is the variation of the effective diffusivity of the rods with space, which is obtained from a least-squares linear fit ($r^2 = 0.84$) to the experimental data in Figure 2g. We solve equations (6) and (7) using a 2$^{\text{nd}}$-order centered-difference method in space and a 1$^{\text{st}}$-order backward Euler method in time. We assume a uniform distribution of rods for the initial condition and impose no-flux boundary conditions on the channel walls. Further details on the numerical method are provided in the SI.

## III. Results and Discussion

Since the rods' velocity increases with $H_2O_2$ concentration[10,13,14,27] and decreases with electrolyte concentration[34,42], the linear static gradients in $H_2O_2$ and KCl concentrations result in a spatial variation in the translational speeds of the rods. Figure 2c shows the pathlines of individual rods in the center channel as a function of position, $x/w$, over 100 seconds. The rods on the left ($x/w < 0.5$, high $H_2O_2$ and low KCl concentration) move faster, resulting in longer pathlines (green) relative to the ones on the right ($x/w > 0.5$, locally low $H_2O_2$ and high KCl concentration) over the same time interval (red). We use the rod pathlines to calculate the MSD, $\Delta L^2$, of the rods for the lifetime of the experiment. Figures 2d, e, and f show the MSD as function of time for several positions, $x/w = 0.25, 0.6, 0.9$. The effective diffusivity of the rods, depicted in figures 2d, e, and f, is estimated here as the slope of MSD vs. time in the linear regime ($t \gg D_{rot}^{-1}$) given as $D_{eff} = \frac{\Delta L^2}{\Delta t}$. On the left side of the channel ($x/w = 0.25$), there is relatively high concentration of $H_2O_2$ and low salt, resulting in an effective diffusivity of 120 times the Brownian diffusivity, as shown in Figure 2d. In the channel center, $x/w = 0.6$, the effective diffusivity decreases to $D_{eff}/D_B = 30.7$ (Figure 2e). On the right-hand side where the salt is highly concentrated and there is little peroxide, $x/w = 0.9$, the effective diffusivity reduces to roughly six times the rods' thermal diffusivity. Figure 2g shows that the effective diffusivity of the rods decreases roughly linearly with space, $x/w$. This gradient arises from the gradient in velocity of the individual rods.

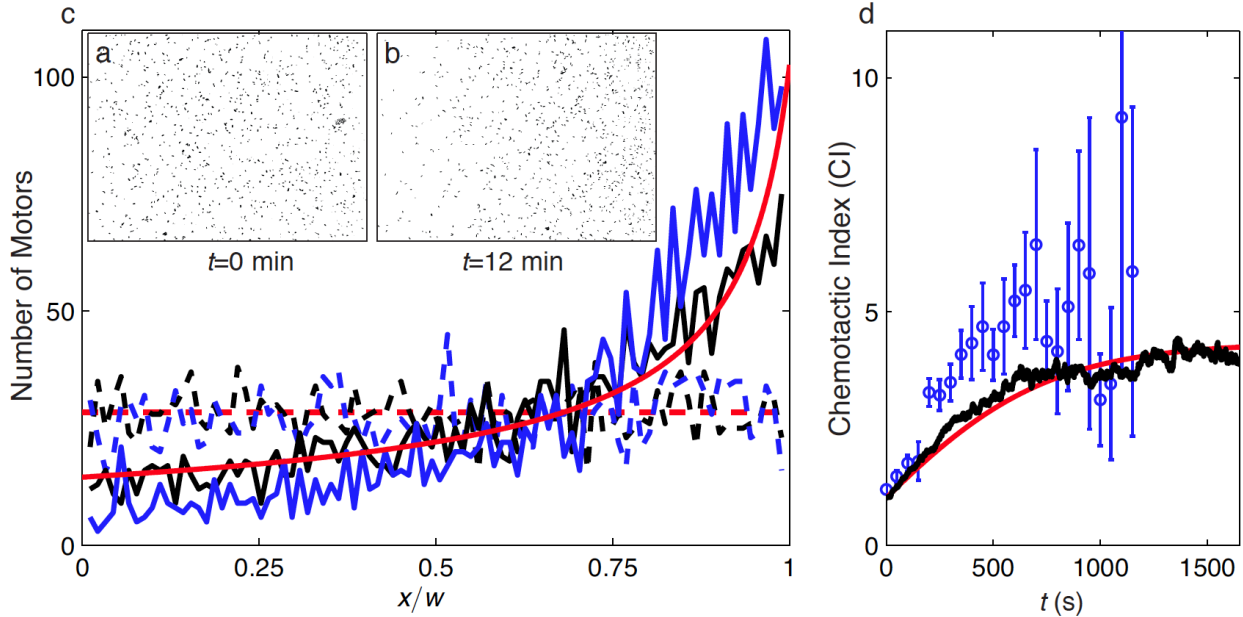

**Figure 3: Chemokinesis-driven accumulation of bimetallic Pt/Au rods in low-mobility regions.** (a) Micrograph showing initial ($t = 0$) uniform distribution of rods in the center channel in the presence of $H_2O_2$ and KCl gradients. (b) Micrograph at $t = 12$ min showing the non-uniform pseudo-equilibrium rod distribution. The rods accumulate on the right side due to a gradient in swimming velocity and effective diffusivity imposed by linear gradients in background $H_2O_2$ and KCl. (c) Number of rods (2550 in total) as a function of space at the beginning of the experiment (dashed lines) and the steady state distribution (solid lines). The number of rods is shown for the experiments (blue), the Brownian Dynamics model (black), and continuum PDE model (red). Panels (a)-(c) appeared in a previous review by our group[28]. (d) Chemotactic index as function of time shown for the experiments (blue symbols), BD model (black), and PDE model (red). The error bars represent one standard deviation. The chemotactic index is initially unity and, for this case, increases until it reaches an equilibrium value of 4.6. Two movies showing accumulation are included in the Electronic Supplementary Information: one showing an experiment and one showing a simulation case.

Figure 3a shows an instantaneous micrograph of the rods in the center channel at $t = 0$, showing a relatively uniform initial distribution of rods. We plot the number of rods as a function of position in Figure 3c for the experiments, simulations, and model. Corresponding to Figure 3a, at $t = 0$ the rods are uniformly dispersed throughout the channel. After 12 min, the rods accumulate on the right due to chemokinesis, as shown in Figure 3b. At this time, the rod number distribution reaches a pseudo-equilibrium state in which the rods are concentrated in the high-salt, low peroxide region on the right as shown in Figure 3b. The temporal evolution of chemokinesis in both simulations and experiments is included as a video in the Supplementary Information (SI). The SI also includes a description of control experiments that showed no migration of rods in a KCl gradient alone (no peroxide), effectively eliminating diffusiophoresis due to the KCl gradient as an alternative mechanism for the rods' accumulation. We plot both the BD and PDE model predictions for the number density as a function of space in Figure 3c, which show good agreement with experimental results. We also note the qualitative resemblance between the quasi-steady-state distributions (solid lines) in Figure 3c and the steady-state probability density function in Figure 2a of Ghosh et al.[55], who also found a roughly inverse relationship between speed and steady-state accumulation.

To quantify the asymmetry in the accumulation of rods at steady state, we define a chemotactic index, CI. Our use of the term "chemotactic index" should not be construed as a claim that the rods exhibit chemotaxis here; rather, we use this term because it has been used in previous literature[79,80] to denote accumulation of microswimmers in a given region. Indeed, in this system accumulation results not from chemotaxis, but from chemokinesis. The CI is defined as the number of rods in the high-salt, low-peroxide concentration region (right hand side) divided by the number of rods in the low-salt, high-peroxide concentration region (left hand side)[81]. Specifically,

$$CI = \frac{N_R}{N_L}, \qquad (8)$$

where $N_R$ and $N_L$ are the number of rods in the rightmost and leftmost regions of the main channel, respectively. The width of the rightmost and leftmost regions is defined to be 1/10th of the channel width. Figure 3d shows the CI as a function of time. The CI is initially unity and increases as $CI(t)/CI_{final} = 1 - e^{-t/\tau}\left(1 + CI_{final}^{-1}\right)$ over $\tau \sim 4$ min until it reaches a pseudo-equilibrium $CI \sim 5$. Cases with large diffusivity gradient, $dD/dx$, and small minimum diffusivity, $D_{min}$, tend to attain higher equilibrium CI values.

Although Figure 3c shows good agreement among the experiments, simulations, and model after 12 minutes, Figure 3d shows some disparity between the experimental estimates of CI and the BD and PDE predictions. Agreement between experiments and predictions may be improved by introducing a weak dependence of rotational diffusivity on position. Specifically, if $D_{rot}$ were to increase with increasing $H_2O_2$ concentration and/or with decreasing KCl concentration, this would effectively increase the predicted steady-state CI. Rods in high-salt, low-H2O2 regions may reorient less often (and thus be less likely to escape the low-motility region, where they accumulate) and rods in high-H2O2, low-salt regions would reorient more often (and thus escape more frequently to low-motility regions). In the PDE model, this would correspond to a case in which $0.5 < \alpha < 1$[74].

Some of the disparity between the experiments and model predictions may also be attributed to hydrodynamic and electrostatic interactions among rods, which were ignored in both the Fokker-Planck and BD models. In the low-speed (salt-rich, peroxide-poor) regions, rods accumulate over time and therefore come close to each other more often. More detailed computational models (beyond the scope of this work) could reveal that the non-spherical nature of the rods, combined with their finite size, could produce hydrodynamic effects in the form of additional self-propulsion velocity terms and additional torques. Additionally, each rod generates its own electric field to propel itself, and so the interactions among the self-generated fields of neighboring rods could also play a role. For simplicity we neglect these complications here and show that while self-polarization may be occurring to a limited extent, it is not necessary to explain the observed accumulation.

To demonstrate the time scale and repeatability of the chemokinesis-based accumulation, we conducted a separate experiment in which the directions of the $H_2O_2$ and KCl gradients were reversed several times, and in response the rods accumulated on alternating sides of the channel. The results are depicted in Figure 4 (note the horizontal *x*-axis and vertical time axis). The channel was subdivided into horizontal bins and the number of rods in each bin was measured at each time. Initially ($t = 0$), the KCl reservoir was at $x = 0$ and the $H_2O_2$ reservoir at $x = w$, so that $H_2O_2$ concentration increased from left to right and KCl concentration increased from right to left. This arrangement resulted in accumulation of the rods near $x = 0$, where the rods' motility was minimized, which is visible after 800 sec. At $t = 800$ sec, the gradients were switched such that the $H_2O_2$ concentration increased from right to left and KCl concentration increased from left to right. Over a 20-minute period, the particles underwent a net chemokinetic accumulation on the opposite side of the channel. At $t = 3100$ seconds and $t = 4800$ seconds, the gradients were switched again, resulting in the same chemokinetic accumulation response in regions of minimal motility. Red dashes indicate the times when the gradient directions were switched. This experiment demonstrates the repeatability of chemokinesis-induced accumulation and shows that the time scale for chemokinesis to occur is on the order of 1500-2000 sec.

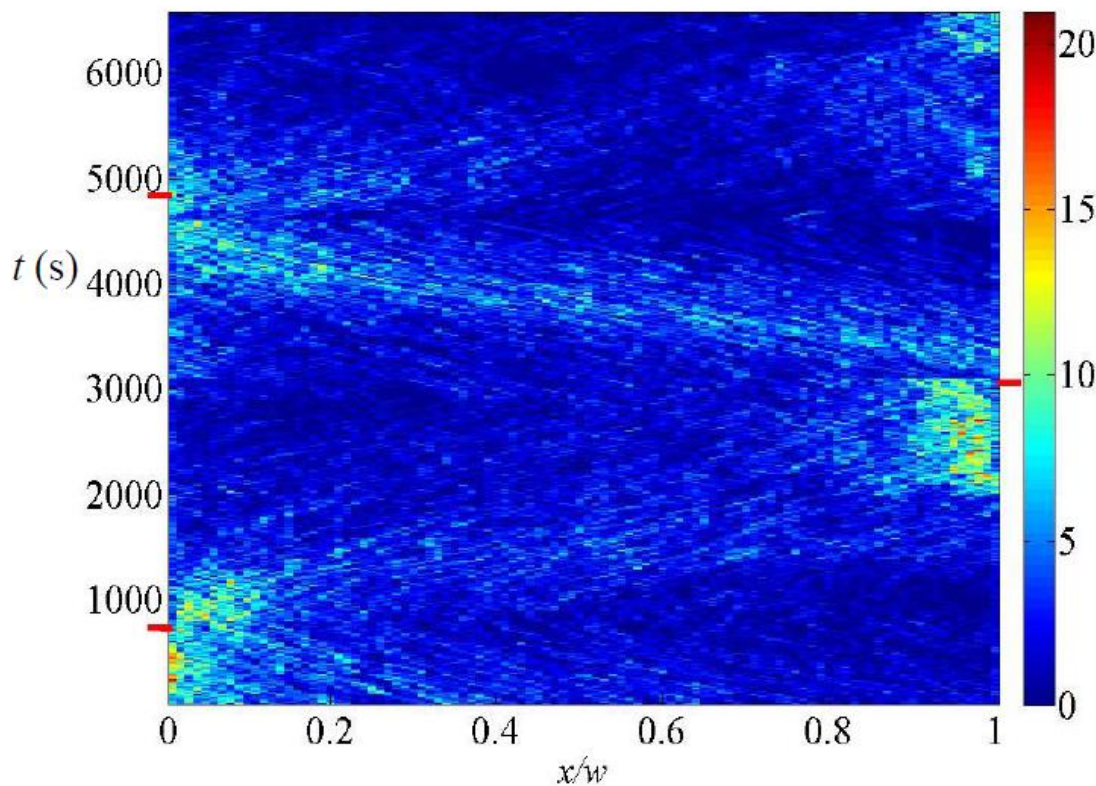

**Figure 4: Accumulation of rods in alternating sides of the microchannel.** In this figure, the channel is divided into 100 bins. The color represents the average number of rods in that bin over a period of 100 seconds. Initially (t = 0), the experiment is set up such that the H2O2 and KCl concentrations increase and decrease from x = 0 to x = w, respectively. The gradient directions are switched at t = 800, 3100, and 4800 sec (indicated by red dashes). In each case, the rods preferentially accumulate in high-KCl, low-H2O2 regions (i.e., where the rods' speed is minimized) within 1000 seconds.

One can use standard diffusion theory to estimate the timescale over which the chemokinesis-based accumulation occurs. For Fickian diffusion in one dimension with diffusivity $D$, the root-mean-square displacement (the square root of the MSD), $L$, scales with the square root of time $t$ as

$$L = \sqrt{2Dt}. \tag{9}$$

The characteristic time for diffusive particles to traverse a distance equal to $L$ is then

$$t = \frac{L^2}{2D}. \tag{10}$$

Let $L$ be the width of the channel, in our case 400 μm. As discussed, the minimum and maximum values of effective diffusivity in the experiments range from roughly $D_{min} = 6D_B$ to $D_{max} = 120D_B$ (see Figure 2g), where $D_B = 1$ μm²/s. Inserting numerical values, we estimate that accumulation should begin to become significant after approximately $t = L^2/2D_{max} = 666$ seconds (about 11 minutes). As shown in Figure 3b and 3c, accumulation is quantifiably present after 12 minutes, in rough agreement with this prediction.

Using the PDE model, we compiled a phase map of the pseudo-equilibrium CI as a function of $D_{min}$ and $dD/dx$. The theoretical phase map along with the experimental results is shown in Figure 5a. The phase map exhibits three distinct regions: (1) at low diffusivity gradient magnitude, there is no net accumulation and the rods exhibit nearly uniform velocity/diffusivities over the entire space; (2) when the effective diffusivity of the rods dips below the Brownian diffusivity $D_{min} < D_B$, this represents the case where the particles aggregate (and stop moving) or reach a sticky boundary; (3) the physical regime where chemokinesis occurs, is that for which $D_{min} > D_B$ and $|dD/dx| > 0$. The experiments are performed for a variety of effective diffusivity gradients $dD/dx$, but all have a minimum effective diffusivity near $D_B$. If we normalize the effective diffusivity gradient by the minimum diffusivity, we can collapse the entire phase map onto a single line as shown in Figure 5b, which compares the predictions of the PDE model (solid line) to experimental measurements (squares) and BD simulations (circles). The experiments, model, and BD simulations show good agreement and indicate that the CI increases with increasing diffusivity gradient. This is expected because stronger gradients result in greater net drift of high-diffusivity particles into regions with low diffusivities.

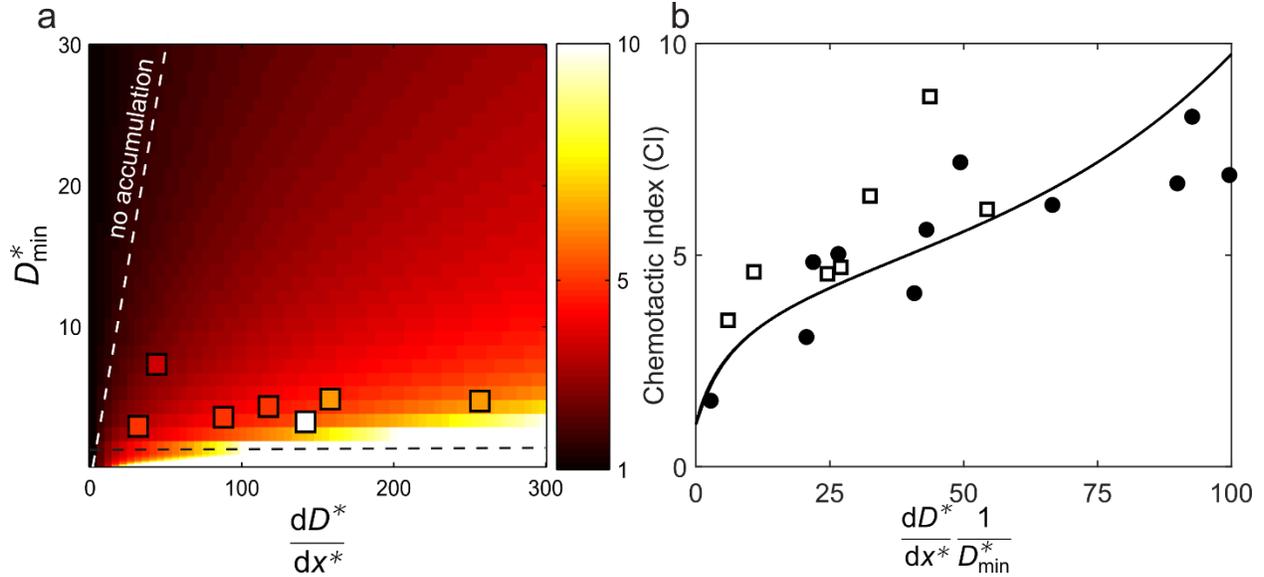

**Figure 5: Steady-state chemotactic index.** (a) Contour map generated from PDE model (equation (1)) showing the equilibrium chemotactic index, *CI*, as function of the normalized effective diffusivity gradient $dD^*/dx^* = d(D_{eff}/D_B)/d(x/w)$ and $D^*_{min} = D_{eff,min}/D_B$. Experiments are shown as colored squares, where the color corresponds to the chemotactic index. The chemotactic index increases with decreasing minimum diffusivity and increasing diffusivity gradient. Larger gradients result in a higher asymmetry in diffusion (higher on left) and smaller minimum diffusivities result in rods that tend to remain at the right end where salt concentrations are large. The white dashed line delineates a region where there is negligible accumulation ($CI \approx 1$) because the gradients in diffusivity are too small to result in asymmetric diffusion. A dashed black line shows the rod's Brownian diffusivity. Large CI values are attained below this diffusivity because the rods effectively become spatially fixed due to their lack of motility. This corresponds to rods aggregating and losing motility or becoming stuck to a physical boundary. (b) For linear effective diffusivity gradients, the contour plot can be collapsed onto to a single line of CI as function of $dD^*/(dx^* D^*_{min})$. Open squares show experimental data, while black dots show results from Brownian Dynamics simulations. For the BD simulation data, the abscissa is varied by changing the minimum and maximum speeds, which in general simultaneously varies the effective diffusivity gradient and the minimum effective diffusivity in the system.

Since the addition of electrolytes increases the viscosity of an aqueous solution, we sought to rule out viscosity as a potential contributor to the accumulation of nanorods in KCl-rich regions of the channel. The viscosity of aqueous KCl solutions was measured and tabulated by Grimes et al.[82] By interpolating this data (presented in the SI), we found that the viscosity of a KCl solution at the maximum concentration considered in our experiments (approximately 54.5 µM) exceeds that of pure water by 0.00014% (less than 1 part per thousand). Thus, we conclude that KCl-induced viscosity increases play a negligible role in the chemokinetic accumulation of rods described in this work.

## IV. Summary

We have shown that Pt/Au self-propelled rods accumulate in low-motility regions when immersed in static fuel gradients because of their chemokinetic response to $H_2O_2$ and salt. This chemokinesis-induced accumulation is similar to that observed in various prokaryotic and eukaryotic cells[20]. We demonstrated good agreement between experiments and both Brownian Dynamics simulations and a conservation equation based on the Fokker-Planck Equation. Chemokinesis-based accumulation has the potential to be leveraged for a variety of applications including environmental remediation, targeted drug delivery, self-healing materials, and more.

Although this study has focused on Pt/Au rods, it is important to note that our findings are expected to apply generally to a variety of self-propelled particles that exhibit an orthokinetic response to fuel concentration. The key components of this system necessary for chemokinesis-driven accumulation, namely orthokinetic response to fuel and random motion trajectories, are observed in a wide variety of active colloid systems. In any active colloid system in which the particle trajectory varies in a stochastic manner, the particles exhibit an orthokinetic response, and the gradient is reasonably strong, one can expect chemokinesis-driven accumulation will be observed and the phase diagram in Figure 5a will be applicable.

In the present work, we have assumed klinokinesis to be negligible. That is, we have assumed the rods' rotational velocity is independent of chemical concentration and is governed exclusively by rotational diffusion. Likewise, the simulations assumed pure orthokinesis, and the agreement between the simulations and experiments appears to validate this assumption. However, there are active colloid systems, such as Quincke rollers[83] or Pt/Au-based microgears[84], in which the rotational speed clearly depends on stimulus intensity (e.g. electric field magnitude or local fuel concentration), and the collective dynamics of these systems are the subject of ongoing study[85]. A logical extension of our work would be to study an active colloid system that exhibits both orthokinesis and klinokinesis, and to elucidate the role of each in determining accumulation of particles in certain locations.

## Author Contributions

J.L.M., P.M.W., and J.D.P. conceived the project. J.L.M. and P.M.W. designed and performed the experiments. P.M.W. and J.L.M. wrote the code for and performed the Brownian Dynamics simulations and PDE model. N.A.M. wrote the video analysis software. J.L.M., P.M.W., and J.D.P. wrote the manuscript, and all authors provided feedback and revisions.

## Acknowledgements

J.L.M. and P.M.W. acknowledge support from National Science Foundation Graduate Research Fellowships. N.A.M. acknowledges support from Science Foundation Arizona. This work received additional support from the George Mason University Department of Mechanical Engineering and NSF grant CBET-0853379 (J.D.P.).

## Competing Interests

The authors declare no competing interest.

## Materials & Correspondence

Correspondence and material requests should be directed to J.L.M. (jmoran23@gmu.edu).